\documentclass{article}

\usepackage{microtype}
\usepackage{graphicx}
\usepackage{subfigure}
\usepackage{booktabs} 

\usepackage{amsmath,amsfonts,bm}

\def\eqref#1{equation~\ref{#1}}

\def\1{\bm{1}}

\DeclareMathAlphabet{\mathsfit}{\encodingdefault}{\sfdefault}{m}{sl}
\SetMathAlphabet{\mathsfit}{bold}{\encodingdefault}{\sfdefault}{bx}{n}

\usepackage{url}
\usepackage{multirow}
\usepackage{color}
\usepackage{xspace}

\newcommand{\BERTforCode}{CuBERT\xspace}

\newcommand{\update}[1]{#1}

\usepackage{hyperref}

\usepackage[group-separator={,},group-minimum-digits = 4]{siunitx}

\usepackage{cleveref}
\usepackage{listings}
\usepackage{color}

\usepackage{authblk}

\definecolor{dkgreen}{rgb}{0,0.6,0}
\definecolor{gray}{rgb}{0.5,0.5,0.5}
\definecolor{mauve}{rgb}{0.58,0,0.82}

\usepackage[accepted]{icml2020}

\icmltitlerunning{Learning and Evaluating Contextual Embedding of Source Code}

\begin{document}

\twocolumn[
\icmltitle{Learning and Evaluating Contextual Embedding of Source Code}



\icmlsetsymbol{equal}{*}

\begin{icmlauthorlist}
\icmlauthor{Aditya Kanade}{equal,iisc,google}
\icmlauthor{Petros Maniatis}{equal,google}
\icmlauthor{Gogul Balakrishnan}{google}
\icmlauthor{Kensen Shi}{google}
\end{icmlauthorlist}

\icmlaffiliation{iisc}{Indian Institute of Science, Bangalore, India}
\icmlaffiliation{google}{Google Brain, Mountain View, USA}

\icmlcorrespondingauthor{Aditya Kanade}{kanade@iisc.ac.in}
\icmlcorrespondingauthor{Petros Maniatis}{maniatis@google.com}

\icmlkeywords{Machine Learning, ICML, Program Understanding, BERT for Code, Program Repair, Pre-trained Code Embedding}

\vskip 0.3in
]



\printAffiliationsAndNotice{\icmlEqualContribution} 

\begin{abstract}
Recent research has achieved impressive results on understanding and improving source code
by building up on machine-learning techniques developed for natural languages.
A significant advancement in natural-language understanding has come with the development of pre-trained contextual embeddings, such as BERT, which can be fine-tuned for downstream tasks with less labeled data and training budget, while achieving better accuracies.
However, there is no attempt yet to obtain a high-quality contextual embedding of source code, and to evaluate it on multiple program-understanding tasks simultaneously; that is the gap that this paper aims to mitigate.
Specifically, first, we curate a massive, deduplicated corpus of 7.4M Python files from GitHub, which we use to pre-train \BERTforCode, an open-sourced code-understanding BERT model;
and, second, we create an open-sourced benchmark that comprises five classification tasks and one program-repair task, akin to code-understanding tasks proposed in the literature before.
We fine-tune \BERTforCode on our benchmark tasks, and compare the resulting models to different variants of Word2Vec token embeddings, BiLSTM and Transformer models, as well as published state-of-the-art models, showing that \BERTforCode outperforms them all, even with shorter training, and with fewer labeled examples.
Future work on source-code embedding can benefit from reusing our benchmark, and from comparing against \BERTforCode models as a strong baseline.
\end{abstract}

\section{Introduction}

Modern software engineering places a high value on writing clean and readable code. This helps other developers understand the author's intent so that they can maintain and extend the code. Developers use meaningful identifier names and natural-language documentation to make this happen~\citep{Martin:2008:CCH:1388398}. As a result, source code contains substantial information that can be exploited by machine-learning algorithms. Indeed, sequence modeling on source code has been shown to be successful in a variety of software-engineering tasks, such as code completion~\citep{6227135,Raychev:2014:CCS:2594291.2594321}, source code to pseudo-code mapping~\citep{oda2015learning}, API-sequence prediction~\citep{Gu:2016:DAL:2950290.2950334}, program repair~\citep{Pu:2016:SNP:2984043.2989222,Gupta:2017:DFC:3298239.3298436}, and natural language to code mapping~\citep{DBLP:conf/emnlp/IyerKCZ18}, among others.

The distributed vector representations of tokens, called token (or word) embeddings, are a crucial component of neural methods for sequence modeling. Learning useful embeddings in a supervised setting with limited data is often difficult. Therefore, many unsupervised learning approaches have been proposed to take advantage of large amounts of unlabeled data that are more readily available. This has resulted in ever more useful pre-trained token embeddings~\citep{DBLP:journals/corr/abs-1301-3781,Pennington14glove:global,bojanowski2017enriching}. However, the subtle differences in the meaning of a token in varying contexts are lost when each word is associated with a single representation. Recent techniques for learning contextual embeddings~\citep{NIPS2017_7209,peters2018deep,GPT,radford2019language,devlin-etal-2019-bert,DBLP:journals/corr/abs-1906-08237} provide ways to compute representations of tokens based on their surrounding context, and have shown significant accuracy improvements in downstream tasks, even with only a small number of task-specific parameters.

Inspired by the success of pre-trained contextual embeddings for natural languages, we present the first attempt to apply the underlying techniques to source code. In particular, BERT~\citep{devlin-etal-2019-bert} produces a bidirectional Transformer encoder~\citep{NIPS2017_7181} by training it to predict values of masked tokens, and whether two sentences follow each other in a natural discourse. The pre-trained model can be fine-tuned for downstream supervised tasks and has been shown to produce state-of-the-art results on a number of natural-language understanding benchmarks. In this work, we derive a contextual embedding of source code by training a BERT model on source code. We call our model \BERTforCode, short for \emph{Code Understanding BERT}. 

In order to achieve this, we curate a massive corpus of Python programs collected from GitHub. GitHub projects are known to contain a large amount of duplicate code. To avoid biasing the model to such duplicated code, we perform deduplication using the method of~\citet{DBLP:journals/corr/abs-1812-06469}. The resulting corpus has \update{\num{7.4}} million files with a total of \update{\num{9.3} billion tokens} (\num{16} million unique). For comparison, we also train Word2Vec embeddings~\citep{DBLP:journals/corr/abs-1301-3781,NIPS2013_5021}, namely, continuous bag-of-words (CBOW) and Skipgram embeddings, on the same corpus.

For evaluating \BERTforCode, we create a benchmark of five classification tasks, and a sixth localization and repair task. The classification tasks range from classification of source code according to presence or absence of certain classes of bugs, to mismatch between a function's natural language description and its body, to predicting the right kind of exception to catch for a given code fragment. The localization and repair task, defined for variable-misuse bugs~\citep{DBLP:journals/corr/abs-1904-01720}, is a pointer-prediction task.
Although similar tasks have appeared in prior work, the associated datasets come from different languages and varied sources; instead we create a cohesive multiple-task benchmark dataset in this work. To produce a high-quality dataset, we ensure that there is no overlap between pre-training and fine-tuning examples, and that all of the tasks are defined on Python code.

We fine-tune \BERTforCode on each of the {classification} tasks and compare the results to multi-layered bidirectional LSTM~\citep{Hochreiter:1997:LSM:1246443.1246450} models, as well as Transformers~\citep{NIPS2017_7181}. We train the LSTM models from scratch and also using pre-trainined Word2Vec embeddings. Our results show that \BERTforCode consistently outperforms these baseline models by \update{\SIrange{3.2}{14.7}{\percent}} across the classification tasks. We perform a number of additional studies by varying the sampling strategies used for training Word2Vec models, and by varying program lengths. In addition, we also show that \BERTforCode can be fine-tuned effectively using only 33\% of the task-specific labeled data and with only 2 epochs, and that, even then, it attains results competitive to the baseline models trained with the full datasets and many more epochs. \BERTforCode, when fine-tuned on the variable-misuse localization and repair task, produces high classification, localization and localization+repair accuracies and outperforms published state-of-the-art models~\cite{hellendoorn2020global,DBLP:journals/corr/abs-1904-01720}.
Our contributions are as follows:
\begin{itemize}
    \item We present the first attempt at pre-training a BERT contextual embedding of source code.
    \item We show the efficacy of the pre-trained contextual embedding on five classification tasks. Our fine-tuned models outperform baseline LSTM models (with/without Word2Vec embeddings), as well as Transformers trained from scratch, even with reduced training data.
    \item We evaluate \BERTforCode on a pointer prediction task and show that it outperforms state-of-the-art results significantly.
    \item We make the models and datasets publicly available.\footnote{\url{https://github.com/google-research/google-research/tree/master/cubert}} We hope that future work benefits from our contributions, by reusing our benchmark tasks, and by comparing against our strong baseline models.
\end{itemize}

\section{Related Work}
\label{sec:related}

Given the abundance of natural-language text, and the relative difficulty of obtaining labeled data, much effort has been devoted to using large corpora to learn about language in an unsupervised fashion, before trying to focus on tasks with small labeled training datasets. Word2Vec~\citep{DBLP:journals/corr/abs-1301-3781,NIPS2013_5021} computed word embeddings based on word co-occurrence and proximity, but the same embedding is used regardless of the context. The continued advances in word~\citep{Pennington14glove:global} and subword~\citep{bojanowski2017enriching} embeddings led to publicly released pre-trained embeddings, used in a variety of tasks.

To deal with varying word context, contextual word embeddings were developed~\citep{NIPS2017_7209,peters2018deep,GPT,radford2019language}, in which an embedding is learned for the \emph{context} of a word in a particular sentence, namely the sequence of words preceding it and possibly following it.
BERT~\citep{devlin-etal-2019-bert} improved natural-language pre-training by using a de-noising autoencoder. Instead of learning a language model, which is inherently sequential, BERT optimizes for predicting a noised word within a sentence. Such prediction instances are generated by choosing a word position and either keeping it unchanged, removing the word, or replacing the word with a random wrong word. It also pre-trains with the objective of predicting whether two sentences can be next to each other. These pre-training objectives, along with the use of a Transformer-based architecture, gave BERT an accuracy boost in a number of NLP tasks over the state-of-the-art. BERT has been improved upon in various ways, including modifying training objectives, utilizing ensembles, combining attention with autoregression~\citep{DBLP:journals/corr/abs-1906-08237}, and expanding pre-training corpora and time~\citep{DBLP:journals/corr/abs-1907-11692}. However, the main architecture of BERT seems to hold up as the state-of-the-art, as of this writing.


In the space of programming languages, embeddings have been learned for specific software-engineering tasks~\citep{chen2019literature}. These include embeddings of variable and method identifiers using local and global context~\citep{Allamanis:2015:SAM:2786805.2786849}, abstract syntax trees (ASTs)~\citep{Mou:2016:CNN:3015812.3016002,zhang2019novel}, AST paths~\citep{Alon:2019:CLD:3302515.3290353}, memory heap graphs~\citep{DBLP:journals/corr/LiTBZ15}, and ASTs enriched with data-flow information~\citep{graphsiclr2018,hellendoorn2020global}. These approaches require analyzing source code beyond simple tokenization. In this work, we derive a pre-trained contextual embedding of tokenized source code without explicitly modeling source-code-specific information, and show that the resulting embedding can be effectively fine-tuned for downstream tasks.

CodeBERT~\citep{feng2020codebert} targets paired natural-language (NL) and multi-lingual programming-language (PL) tasks, such as code search and generation of code documentation. It pre-trains a Transformer encoder by treating a natural-language description of a function and its body as separate sentences in the sentence-pair representation of BERT. We also handle natural language directly, but do not require such a separation. Natural-language tokens can be mixed with source-code tokens both within and across sentences in our encoding. One of our benchmark tasks, function-docstring mismatch, illustrates the ability of \BERTforCode to handle NL-PL tasks.

\section{Experimental Setup}

We now outline our benchmarks and experimental study. The supplementary material contains deeper detail aimed at reproducing our results.

\subsection{Code Corpus for Fine-Tuning Tasks}
\label{sec:py150}

We use the ETH Py150 corpus~\citep{ethpy150} to generate datasets for the fine-tuning tasks. This corpus consists of \update{150K} Python files from GitHub, and is partitioned into a training split (\update{100K} files) and a test split (\update{50K} files). We held out \update{10K} files from the training split as a validation split.
We deduplicated the dataset in the fashion of \citet{DBLP:journals/corr/abs-1812-06469}.
Finally, we drop from this corpus those projects for which licensing information was not available or whose licenses restrict use or redistribution. We call the resulting corpus the \emph{ETH Py150 Open} corpus.\footnote{\url{https://github.com/google-research-datasets/eth_py150_open}}
This is our Python fine-tuning code corpus, and it consists of \num{74749} training files, \num{8302} validation files, and \num{41457} test files.

\subsection{The GitHub Python Pre-Training Code Corpus}
\label{sec:github}

We used the public GitHub repository hosted on Google's BigQuery platform (the \texttt{github\_repos} dataset under BigQuery's public-data project, \texttt{bigquery-public-data}). We extracted all files ending in \texttt{.py}, under open-source, redistributable licenses, removed symbolic links, and retained only files reported to be in the \texttt{refs/heads/master} branch. This resulted in about \update{\num{16.2} million} files.

To avoid duplication between pre-training and fine-tuning data, we removed files that had high similarity to the files in the ETH Py150 Open corpus, using the method of \citet{DBLP:journals/corr/abs-1812-06469}. In particular, two files are considered similar to each other if the Jaccard similarity between the sets of tokens (identifiers and string literals) is above 0.8 and in addition, it is above 0.7 for multi-sets of tokens. This brought the dataset to \update{\num{14.3} million} files. We then further deduplicated the remaining files, by clustering them into equivalence classes holding similar files according to the same similarity metric, and keeping only one exemplar per equivalence class. This helps avoid biasing the pre-trained embedding. Finally, we removed files that could not be parsed.
In the end, we were left with \update{\num{7.4}} million Python files containing over \update{\num{9.3} billion tokens}. This is our Python pre-training code corpus.

\subsection{Source-Code Modeling}
\label{sec:source-code-modeling}

We first tokenize a Python program using the standard Python tokenizer (the \texttt{tokenize} package). We leave language keywords intact and produce special tokens for syntactic elements that have either no string representation (e.g., \texttt{DEDENT} tokens, which occur when a nested program scope concludes), or ambiguous interpretation (e.g., new-line characters inside string literals, at the logical end of a Python statement, or in the middle of a Python statement result in distinct special tokens). We split identifiers according to common heuristic rules (e.g., snake or Camel case). Finally, we split string literals using heuristic rules, on white-space characters, and on special characters. \update{We limit all thus produced tokens to a maximum length of 15 characters.} We call this the \emph{program vocabulary}.
Our Python pre-training code corpus contained \update{\num{16} million unique tokens}.

We greedily compress the program vocabulary into a \emph{subword vocabulary}~\citep{37842} using the \texttt{SubwordTextEncoder} from the Tensor2Tensor project~\citep{DBLP:conf/amta/VaswaniBBCGGJKK18}\footnote{\url{https://github.com/tensorflow/tensor2tensor/blob/master/tensor2tensor/data_generators/text_encoder.py}}, resulting in \update{about 50K tokens}. All words in the program vocabulary can be losslessly encoded using one or more of the subword tokens.

We tokenize programs first into program tokens, as described above, and then encode those tokens one by one in the subword vocabulary. The objective of this encoding scheme is to preserve syntactically meaningful boundaries of tokens. For example, the identifier ``\texttt{snake\_case}'' could be encoded as \mbox{``\texttt{sna ke \_ ca se}''}, preserving the snake case split of its characters, even if the subtoken ``\texttt{e\_c}'' were very popular in the corpus; the latter encoding might result in a smaller representation but would lose the intent of the programmer in using a snake-case identifier. Similarly, ``\texttt{i=0}'' may be very frequent in the corpus, but we still force it to be encoded as separate tokens \texttt{i}, \texttt{=}, and \texttt{0}, ensuring that we preserve the distinction between operators and operands. Both the BERT model and the Word2Vec embeddings are built on the subword vocabulary.

\subsection{Fine-Tuning Tasks}
\label{sec:fine-tuningTasks}

To evaluate \BERTforCode, we design five {classification tasks and a multi-headed pointer task}. These are motivated by prior work, but unfortunately, the associated datasets come from different languages and varied sources. We want the tasks to be on Python code, and for accurate results, we ensure that there is no overlap between pre-training and fine-tuning datasets. We therefore create all the tasks on the ETH Py150 Open corpus (see Section~\ref{sec:py150}). As discussed in Section~\ref{sec:github}, we ensure that there is no duplication between this and the pre-training corpus. We hope that our datasets for these tasks will be useful to others as well. The fine-tuning tasks are described below. \update{A more detailed discussion is presented in the supplementary material.}

\begin{table*}[t]
\centering
\begin{tabular}{lrrr}
\toprule
          & Train & Validation & Test \\ \midrule
 Variable-Misuse Classification & \num{700708} & \num{8192} \;(\num{75478}) & \num{378440} \\
 Wrong Binary Operator & \num{459400} & \num{8192} \;(\num{49804}) & \num{251804} \\
 Swapped Operand & \num{236246} & \num{8192} \;(\num{26118}) & \num{130972} \\
 Function-Docstring & \num{340846} & \num{8192} \;(\num{37592}) & \num{186698} \\
 Exception Type & \num{18480} & \num{2088}\phantom{0} \;(\num{2088}) & \num{10348}\\
 Variable-Misuse Localization and Repair  & \num{700708} & \num{8192} \;(\num{75478}) & \num{378440} \\
\bottomrule
\end{tabular}
\caption{\label{tab:BenchmarkDatasets} \update{Benchmark fine-tuning datasets. Note that for validation, we have subsampled the original datasets (in parentheses) down to \num{8192} examples, except for exception classification, which only had \num{2088} validation examples, all of which are included.}}
\label{tab:datasets}
\end{table*}

\paragraph{Variable-Misuse {Classification}}
\citet{graphsiclr2018} observed that developers may mistakenly use an incorrect variable in the place of a correct one. These mistakes may occur when developers copy-paste similar code but forget to rename all occurrences of variables from the original fragment, or when there are similar variable names that can be confused with each other. These can be subtle errors that remain undetected during compilation. The task by \citet{graphsiclr2018} is to choose the correct variable name at a location within a C\# function. We take the classification version restated by \citet{DBLP:journals/corr/abs-1904-01720}, wherein, given a function, the task is to predict whether there is a variable misuse at \emph{any} location in the function, without specifying a particular location to consider. Here, the classifier has to consider all variables and their usages to make the decision. In order to create negative (buggy) examples, we replace a variable use at some location with another variable that is defined within the function.

\paragraph{Wrong Binary Operator}
\citet{Pradel:2018:DLA:3288538.3276517} proposed the task of detecting whether a binary operator in a given expression is correct. They use features extracted from limited surrounding context. We use the entire function with the goal of detecting whether any binary operator in the function is incorrect. The negative examples are created by randomly replacing some binary operator with another type-compatible operator.

\paragraph{Swapped Operand}
\citet{Pradel:2018:DLA:3288538.3276517} propose the wrong binary operand task where a variable or constant is used incorrectly in an expression, but that task is quite similar to the variable-misuse task we already use.
We therefore define another class of operand errors where the operands of non-commutative binary operators are swapped. The operands can be arbitrary subexpressions, and are not restricted to be just variables or constants.  To simplify example generation, we restrict this task to examples in which the operator and operands all fit within a single line. 

\paragraph{Function-Docstring Mismatch}
Developers are encouraged to write descriptive docstrings to explain the functionality and usage of functions. This provides parallel corpora between code and natural language sentences that have been used for machine translation~\citep{barone2017parallel}, detecting uninformative docstrings~\citep{louis2018deep} and to evaluate their utility to provide supervision in neural code search~\citep{cambronero2019deep}. We prepare a sentence-pair classification problem where the function and its docstring form two distinct sentences. The positive examples come from the correct function-docstring pairs. We create negative examples by replacing correct docstrings with docstrings of other functions, randomly chosen from the dataset. For this task, the existing docstring is removed from the function body. 

\paragraph{Exception Type}
While it is possible to write generic exception handlers (e.g., ``\texttt{except Exception}'' in Python), it is considered a good coding practice to catch and handle the precise exceptions that can be raised by a code fragment.\footnote{\url{https://google.github.io/styleguide/pyguide.html##24-exceptions}} We identified the 20 most common exception types from the GitHub dataset, excluding the catch-all \texttt{Exception} (\update{full list in Table~1 in the supplementary material}). Given a function with an \texttt{except} clause for one of these exception types, we replace the exception with a special ``hole'' token. The task is the multi-class classification problem of predicting the original exception type.

\paragraph{{Variable-Misuse Localization and Repair}}
As an instance of a non-classification task, we consider the joint classification, localization, and repair version of the variable-misuse task from~\citet{DBLP:journals/corr/abs-1904-01720}. Given a function, the task is to predict one pointer (called the localization pointer) to identify a variable-misuse location, and another pointer (called the repair pointer) to identify a variable from the same function that is the right one to use at the faulty location. The model is also trained to classify functions that do not contain any variable misuse as bug-free by making the localization pointer point to a special location in the function. We create negative examples using the same method as used in the Variable-Misuse Classification task.

Table~\ref{tab:BenchmarkDatasets} lists the sizes of the resulting benchmark datasets extracted from the fine-tuning corpus. The Exception Type task contains significantly fewer examples than the other tasks, since examples for this task only come from functions that catch one of the chosen 20 exception types.

\subsection{BERT for Source Code}
\label{sec:bert-for-code}

The BERT model~\citep{devlin-etal-2019-bert} consists of a multi-layered Transformer encoder. It is trained with two tasks: (1) to predict the correct tokens in a fraction of all positions, some of which have been replaced with incorrect tokens or the special \texttt{[MASK]} token (the Masked Language Model task, or \emph{MLM}) and (2) to predict whether the two sentences separated by the special \texttt{[SEP]} token follow each other in some natural discourse (the Next-Sentence Prediction task, or \emph{NSP}). Thus, each example consists of one or two \emph{sentences}, where a sentence is the concatenation of contiguous lines from the source corpus, sized to fit the target example length. To ensure that every sentence is treated in multiple instances of both MLM and NSP, BERT by default duplicates the corpus \num{10} times, and generates independently derived examples from each duplicate. With \SI{50}{\percent} probability, the second example sentence comes from a random document (for NSP). \update{A token is chosen at random for an MLM prediction} (up to \num{20} per example), and from those chosen, \SI{80}{\percent} are masked, \SI{10}{\percent} are left undisturbed, and \SI{10}{\percent} are replaced with a random token.

\BERTforCode is similarly formulated, but a \BERTforCode \update{line} is a logical code line, as defined by the Python standard. Intuitively, a logical code line is the shortest sequence of consecutive lines that constitutes a legal statement, e.g., it has correctly matching parentheses. We count example lengths by counting the subword tokens of both sentences (see Section~\ref{sec:source-code-modeling}). 

We train the BERT Large model having 24 layers with 16 attention heads and 1024 hidden units. Sentences are created from our pre-training dataset. Task-specific classifiers pass the embedding of a special start-of-example \texttt{[CLS]} token through feed-forward and softmax layers. For the pointer prediction task, the pointers are computed exactly as by~\citet{DBLP:journals/corr/abs-1904-01720}; whereas in that work, the pointers are computed from the output of an LSTM layer, in our model, they are computed from the last-layer hiddens of  BERT.

\subsection{Baselines}

\subsubsection{Word2Vec}
We train Word2Vec models using the same pre-training corpus as the BERT model. To maintain parity, we generate the dataset for Word2Vec using the same pipeline as BERT but by disabling masking and generation of negative examples for NSP. The dataset is generated without any duplication.
%
We train both CBOW and Skipgram models using GenSim~\citep{rehurek_lrec}. To deal with the large vocabulary, we use negative sampling and hierarchical softmax~\citep{DBLP:journals/corr/abs-1301-3781,NIPS2013_5021} to train the two versions. In all, we obtain four types of Word2Vec embeddings.

\subsubsection{Bidirectional LSTM and Transformer}
In order to obtain context-sensitive encodings of input sequences for the fine-tuning tasks, we use multi-layered bidirectional LSTMs~\citep{Hochreiter:1997:LSM:1246443.1246450} (BiLSTMs). These are initialized with the pre-trained Word2Vec embeddings. To further evaluate whether LSTMs alone are sufficient without pre-training, we also train BiLSTMs with an embedding matrix that is initialized from scratch with Xavier initialization~\citep{glorot2010understanding}. We also trained Transformer models~\citep{NIPS2017_7181} for our fine-tuning tasks. 
We used BERT's own Transformer implementation, to ensure comparability of results. 
{For comparison with prior work, we use the unidirectional LSTM and pointer model from~\citet{DBLP:journals/corr/abs-1904-01720} for the Variable-Misuse Localization and Repair task.}

\section{Experimental Results}

\subsection{Training Details}
\label{sec:training-details}

\BERTforCode's dataset generation duplicates the corpus \num{10} times, whereas Word2Vec is trained without duplication. To compensate for this difference, we trained Word2Vec for \num{10} epochs and \BERTforCode for \num{1} epoch. We chose models by validation accuracy, both during hyperparameter searches, and during model selection within an experiment.

We pre-train \BERTforCode with the default configuration of the BERT Large model, one model per example length (\num{128}, \num{256}, \num{512}, and \num{1024} subword tokens) with batch sizes of \num{8192}, \num{4096}, \num{2048}, and \num{1024} respectively, and the default BERT learning rate of \num{1e-4}. Fine-tuned models also used the same batch sizes as for pre-training, and BERT's default learning rate (\num{5e-5}). For both, we gradually warm up the learning rate for the first \SI{10}{\percent} of examples, which is BERT's default value.

For Word2Vec, when training with negative samples, we choose \update{\num{5}} negative samples. The embedding size for all the Word2Vec pre-trained models is set at \num{1024}.
For the baseline BiLSTM models, we performed a hyperparameter search on each task and pre-training configuration separately (\num{5} tasks, each trained with the four Word2Vec embeddings, plus the randomly initialized embeddings), for the \num{512} example length. For each of these \num{25} task configurations, we varied the number of layers (\numrange{1}{3}), the number of hidden units (\numlist{128;256;512}), the LSTM output dropout probability (\numlist{0.1;0.5}),  and the learning rate (\numlist{1e-3;1e-4;1e-5}). We used the Adam~\citep{kingma2014adam} optimizer throughout, and batch size \num{8192} for all tasks except the Exception-Type task, for which we used batch size \num{64}. \update{Invariably, the best hyperparameter selection had \num{512} hidden units per layer and learning rate of \num{1e-3}, but the number of layers (mostly \num{2} or \num{3}) and dropout probability varied across best task configurations.} 
Though no single Word2Vec configuration is the best, CBOW trained with negative sampling gives the most consistent results overall.

For the baseline Transformer models, we originally attempted to train a model of the same configuration as \BERTforCode. However, the sizes of our fine-tuning datasets seemed too small to train that large a Transformer. Instead, we performed a hyperparameter search for each task individually, for the \num{512} example length. We varied the number of transformer layers (\numrange{1}{6}), hidden units (\numlist{128;256;512}), learning rates (\numlist{1e-3;5e-4;1e-4;5e-5;1e-5}) and batch sizes (\numlist{512;1024;2048;4096}). The best architecture varied across the tasks: for example, \num{5} layers with \num{128} hiddens and the highest learning rate worked best for the Function-Docstring task, whereas for the Exception-Type task, \num{2} layers, \num{512} hiddens, and the second lowest learning rate worked best.

Finally, for our baseline pointer model (referred to as LSTM+pointer below) we searched over the following hyperparameter choices: hidden sizes of \numlist{512;1024}, token embedding sizes of \numlist{512;1024}, and learning rates of \numlist{1e-1;1e-2;1e-3}. We used the Adam optimizer, a batch size of \num{256}, and example length \num{512}. In contrast to the original work~\citep{DBLP:journals/corr/abs-1904-01720}, we generated one pair of buggy/bug-free examples per function (rather than one per variable use, per function, which would bias towards longer functions), and use \BERTforCode's subword-tokenized vocabulary of 50K subtokens (rather than a limited full-token vocabulary, which leaves many tokens out of vocabulary).

\update{We used TPUs for training our models, except for pre-training Word2Vec embeddings, and the pointer model by \citet{DBLP:journals/corr/abs-1904-01720}. For the rest, and for all evaluations, we used P100 or V100 GPUs. All experiments using pre-trained word or contextual embeddings continued to fine-tune weights throughout training.}

\subsection{Research Questions}
We set out to answer the following research questions. We will address each with our results.
\begin{enumerate}
    \item Do contextual embeddings help with source-code analysis tasks, when pre-trained on an unlabeled code corpus? We compare \BERTforCode to BiLSTM models with and without pre-trained Word2Vec embeddings {on the classification tasks} (Section~\ref{sec:BERTforCodeVsWord2Vec}).
    \item Does fine-tuning actually help, or is the Transformer model by itself sufficient? We compare fine-tuned \BERTforCode models to Transformer-based models trained from scratch {on the classification tasks} (Section~\ref{sec:BERTforCodeVsTransformer}).
    \item How does the performance of \BERTforCode on {the classification} tasks scale with the amount of labeled training data? We compare the performance of fine-tuned \BERTforCode models when fine-tuning with \SIlist{33;66;100}{\percent} of the task training data (Section~\ref{sec:fine-tuningWithLessData}).
    \item How does context size affect \BERTforCode? We compare fine-tuning performance for different example lengths {on the classification tasks} (Section~\ref{sec:SmallerLengths}).
    \item How does \BERTforCode perform on complex tasks, against state-of-the-art methods? We implemented and fine-tuned a model for a multi-headed pointer prediction task, namely, the Variable-Misuse Localization and Repair task (Section~\ref{sec:varmisuse-pointer}). We compare it to the models from~\cite{DBLP:journals/corr/abs-1904-01720} and \cite{hellendoorn2020global}.
\end{enumerate}
Except for Section~\ref{sec:SmallerLengths}, all the results are presented for sequences of length \num{512}.
\update{We give examples of classification instances in the supplementary material and include visualizations of attention weights for them.}

\begin{table*}
    \centering
\sisetup{
round-mode = places,
round-precision = 2,
detect-weight = true,
detect-family = true
}%
\begin{tabular}{lll|ccccc}
\toprule
 & \multicolumn{2}{c|}{\textbf{Setting}}             & \textbf{Misuse} & \textbf{Operator} & \textbf{Operand} & \textbf{Docstring} & \textbf{Exception} \\ \midrule 
\multicolumn{1}{c}{\multirow{4}{*}{\textbf{BiLSTM}}} & \multicolumn{2}{l|}{From scratch} &
\SI{76.2927}{\percent}&	\SI{83.648163}{\percent}&	\SI{88.07047}{\percent}&	\SI{76.010776}{\percent}&	\SI{52.78638}{\percent}
\\ \cline{2-3}
& \multirow{2}{*}{CBOW}     & ns &
\SI{80.32751}{\percent}&	\textbf{\SI{86.81924}{\percent}}&	\SI{89.797926}{\percent}&	\textbf{\SI{89.075357}{\percent}}&	\textbf{\SI{67.008513}{\percent}}
\\
\multirow{2}{*}{(\num{100} epochs)}                                                       &                                    & hs &
\SI{78.00424}{\percent}&	\SI{85.84694}{\percent}&	\textbf{\SI{90.143114}{\percent}}&	\SI{87.68962}{\percent}&	\SI{60.31347}{\percent}
\\ \cline{2-3} 
                                                       & \multirow{2}{*}{Skipgram} & ns &
\SI{77.061135}{\percent}&	\SI{85.140365}{\percent}&	\SI{89.31146}{\percent}&	\SI{83.81203}{\percent}&	\SI{60.071594}{\percent}
\\ 
                                                       &                                    & hs &
\textbf{\SI{80.52728}{\percent}}&	\SI{86.33786}{\percent}&	\SI{89.75287}{\percent}&	\SI{88.79628}{\percent}&	\SI{65.06385}{\percent}
\\ \midrule
\multicolumn{1}{c}{\multirow{3}{*}{\textbf{\BERTforCode}}}                                 & \multicolumn{2}{c|}{\num{2} epochs}                           & 
\SI{94.042}{\percent}&	\SI{89.89658}{\percent}&	\SI{92.198956}{\percent}&	\SI{97.20764}{\percent}&	\SI{61.039084}{\percent}
\\
    & \multicolumn{2}{c|}{\num{10} epochs}                          &
\SI{95.13968}{\percent}&	\SI{92.150164}{\percent}&	\SI{93.622464}{\percent}&	\SI{98.07754}{\percent}&	\SI{77.9702}{\percent}
\\
    & \multicolumn{2}{c|}{\num{20} epochs}                          &
\SI{95.213145}{\percent}&	\SI{92.46354}{\percent}&	\SI{93.35517}{\percent}&	\SI{98.08504}{\percent}&	\SI{79.12152}{\percent}
\\ \midrule
    \textbf{Transformer} & \multicolumn{2}{c|}{\num{100} epochs} &
\SI{78.28434}{\percent}&	\SI{76.554555}{\percent}&	\SI{87.82762}{\percent}&	\SI{91.017634}{\percent}&	\SI{49.56463}{\percent}
\\
    \bottomrule
\end{tabular}

\caption{\label{tab:BERTforCode512} \update{Test accuracies of fine-tuned \BERTforCode against BiLSTM (with and without Word2Vec embeddings) and Transformer trained from scratch {on the classification tasks}.} ``ns" and ``hs" respectively refer to negative sampling and hierarchical softmax settings used for training CBOW and Skipgram models. ``From scratch" refers to training with freshly initialized token embeddings, without pre-training.}
\end{table*}

\subsection{Contextual vs.\ Word Embeddings}
\label{sec:BERTforCodeVsWord2Vec}
The purpose of this analysis is to understand how much pre-trained contextual embeddings help, compared to word embeddings.
For each {classification} task, we trained BiLSTM models starting with each of the Word2Vec embeddings, namely, continuous bag of words (CBOW) and Skipgram trained with negative sampling or hierarchical softmax.
We trained the BiLSTM models for \num{100} epochs and the \BERTforCode models for \num{20} epochs,
and all models stopped improving by the end.

The resulting test accuracies are shown in Table~\ref{tab:BERTforCode512} (first 5 rows and next-to-last row). \BERTforCode consistently outperforms BiLSTM (with the best task-wise Word2Vec configuration) on all tasks, by a margin of \update{\SIrange{3.2}{14.7}{\percent}}. Thus, the pre-trained contextual embedding provides superior results even with a smaller budget of \num{20} epochs, compared to the \num{100} epochs used for BiLSTMs.
The Exception-Type classification task has an order of magnitude less training data than the other tasks (see Table~\ref{tab:BenchmarkDatasets}). The difference between the performance of BiLSTM and \BERTforCode is substantially higher for this task. Thus, fine-tuning is of much value for tasks with limited labeled training data.


We analyzed the performance of \BERTforCode with the reduced fine-tuning budget of only 2 and 10 epochs (see the remaining rows of the \BERTforCode section in Table~\ref{tab:BERTforCode512}). \update{Except for the Exception Type task, \BERTforCode outperforms the best 100-epoch BiLSTM within 2 fine-tuning epochs. On the Exception-Type task, \BERTforCode with 2 fine-tuning epochs outperforms all but two configurations of the BiLSTM baseline.
This shows that, even when restricted to just a few fine-tuning epochs, \BERTforCode can reach accuracies that are comparable to or better than those of BiLSTMs trained with Word2Vec embeddings.}

To sanity-check our findings about BiLSTMs, we also trained the BiLSTM models from scratch, without pre-trained embeddings.
The results are shown in the first row of Table~\ref{tab:BERTforCode512}. \update{Compared to those, the use of Word2Vec embeddings performs better by a margin of \SIrange{2.7}{14.2}{\percent}.}

\subsection{Is Transformer All You Need?}
\label{sec:BERTforCodeVsTransformer}

One may wonder if \BERTforCode's promising results derive more from using a Transformer-based model for its classification tasks, and less from the actual, unsupervised pre-training.
Here we compare our results {on the classification tasks} to a Transformer-based model trained from scratch, i.e., without the benefit of a pre-trained embedding.
As discussed in Section~\ref{sec:training-details}, the size of the training data limited us to try out Transformers that were substantially smaller than the CuBERT model (BERT Large architecture).
All the Transformer models were trained for \num{100} epochs during which their performance stopped improving.
We selected the best model within the chosen hyperparameters for each task based on best validation accuracy.

As seen from the last row of Table~\ref{tab:BERTforCode512}, the performance of \BERTforCode is substantially higher than the Transformer models trained from scratch. \update{Thus, for the same choice of architecture (i.e., Transformer) pre-training seems to help by enabling training of a larger and better model.}


\begin{table*}
\centering
\sisetup{
round-mode = places,
round-precision = 2
}%
\begin{tabular}{cc|@{\quad}SSSSS}
\toprule
\textbf{\begin{tabular}[c]{@{}c@{}}Best of\\ \# Epochs\end{tabular}} & \textbf{\begin{tabular}[c]{@{}c@{}}Train\\ Fraction\end{tabular}} & \textbf{Misuse} & \textbf{Operator} & \textbf{Operand} & \textbf{Docstring} & \textbf{Exception} \\ \midrule
\multirow{3}{*}{\textbf{\num{2}}}                      &
\SI{100}{\percent}                                                                                  &
\SI{94.042}{\percent}&	\SI{89.89658}{\percent}&	\SI{92.198956}{\percent}&	\SI{97.20764}{\percent}&	\SI{61.039084}{\percent}
\\
                                                                                           &
\SI{66}{\percent}                                                                                   &
\SI{93.114483}{\percent}&	\SI{88.761044}{\percent}&	\SI{91.61321}{\percent}&	\SI{97.03784}{\percent}&	\SI{19.485295}{\percent}
\\
                                                                                           &
\SI{33}{\percent}                                                                                   &
\SI{91.39687}{\percent}&	\SI{86.422855}{\percent}&	\SI{90.51885}{\percent}&	\SI{96.37577}{\percent}&	\SI{20.08514}{\percent}
\\ \midrule
\multirow{3}{*}{\textbf{\num{10}}}                     &
\SI{100}{\percent}                                                                                  &
\SI{95.13968}{\percent}&	\SI{92.150164}{\percent}&	\SI{93.622464}{\percent}&	\SI{98.07754}{\percent}&	\SI{77.9702}{\percent}
\\
                                                                                           &
\SI{66}{\percent}                                                                                   &
\SI{94.781625}{\percent}&	\SI{91.51269}{\percent}&	\SI{93.36663}{\percent}&	\SI{97.930235}{\percent}&	\SI{75.241876}{\percent}
\\
                                                                                           &
\SI{33}{\percent}                                                                                   &
\SI{94.28061}{\percent}&	\SI{90.66353}{\percent}&	\SI{92.583853}{\percent}&	\SI{97.3587}{\percent}&	\SI{67.33746}{\percent}
\\ \midrule
\multirow{3}{*}{\textbf{\num{20}}}                     &
\SI{100}{\percent}                                                                                  &
\SI{95.213145}{\percent}&	\SI{92.46354}{\percent}&	\SI{93.35517}{\percent}&	\SI{98.08504}{\percent}&	\SI{79.12152}{\percent}
\\
                                                                                           &
\SI{66}{\percent}                                                                                   &
\SI{94.899744}{\percent}&	\SI{91.794693}{\percent}&	\SI{93.39259}{\percent}&	\SI{97.98755}{\percent}&	\SI{77.31231}{\percent}
\\
                                                                                           &
\SI{33}{\percent}                                                                                   &
\SI{94.45026}{\percent}&	\SI{91.08732}{\percent}&	\SI{92.82212}{\percent}&	\SI{97.627056}{\percent}&	\SI{74.98065}{\percent}
\\
\bottomrule
\end{tabular}
\caption{\update{Effects of reducing training-split size on fine-tuning performance on the classification tasks}.}
\label{tab:BERTforCodeWithLessData}
\end{table*}

\subsection{The Effects of Little Supervision}
\label{sec:fine-tuningWithLessData}

The big draw of unsupervised pre-training followed by fine-tuning is that some tasks have small labeled datasets.
We study here how \BERTforCode fares with reduced training data.
We sampled uniformly the fine-tuning dataset to \SIlist{33;66}{\percent} of its size, and produced corresponding training datasets for each classification task. We then fine-tuned the pre-trained \BERTforCode model with each of the \num{3} different training splits. Validation and testing were done with the same original datasets.  Table~\ref{tab:BERTforCodeWithLessData} shows the results.

\begin{table*}
\centering
\sisetup{
round-mode = places,
round-precision = 2
}%
\begin{tabular}{cccccc|c}
\toprule \textbf{Length} & \textbf{Misuse} & \textbf{Operator} & \textbf{Operand} & \textbf{Docstring} & \textbf{Exception} & \textbf{Misuse on BiLSTM}\\ \midrule
128                                   &
\SI{83.96647}{\percent}&	\SI{79.28913}{\percent}&	\SI{78.01503}{\percent}&	\SI{98.188955}{\percent}&	\SI{62.02593}{\percent}&	\SI{74.316657}{\percent}
\\
256                                   & 
\SI{92.02023}{\percent}&	\SI{88.1895}{\percent}&	\SI{88.02847}{\percent}&	\SI{98.135394}{\percent}&	\SI{72.803795}{\percent}&	\SI{78.46932}{\percent}
\\
512                                   &
\SI{95.213145}{\percent}&	\SI{92.46354}{\percent}&	\SI{93.35517}{\percent}&	\SI{98.08504}{\percent}&	\SI{79.12152}{\percent}&	\SI{80.32751}{\percent}
\\
1024                                   &
\SI{95.834655}{\percent}&	\SI{93.37864}{\percent}&	\SI{95.6195}{\percent}&	\SI{97.899705}{\percent}&	\SI{81.26935}{\percent}&	\SI{81.91828}{\percent}
\\
\bottomrule
\end{tabular}
\caption{\update{Best out of 20 epochs of fine-tuning, for four example lengths, on the classification tasks. For contrast, we also include results for Variable Misuse using the BiLSTM Word2Vec (CBOW + ns) classifier as length varies.}}
\label{tab:SmallerLengths}
\end{table*}

\update{The Function Docstring task seems robust to the reduction of the training dataset, both early and late in the fine-tuning process (that is, within \num{2} vs.~\num{20} epochs), whereas the Exception Classification task is heavily impacted by the dataset reduction, given that it has relatively few training examples to begin with. Interestingly enough, for some tasks, even fine-tuning for only 2 epochs and only using a third of the training data outperforms the baselines. For example, for Variable Misuse and Function Docstring, \BERTforCode at 2 epochs and \SI{33}{\percent} of training data substantially outperforms the BiLSTM with Word2Vec and the Transformer baselines.}

\begin{table*}
    \centering
\sisetup{
round-mode = places,
round-precision = 2
}%
    \begin{tabular}{c|c|c|c|c|c|c}
    \toprule \textbf{Model} & \textbf{Test Data} & \textbf{Setting} & \textbf{True} & \textbf{Classification} & \textbf{Localization} & \textbf{Loc+Repair}\\
&    & & \textbf{Positive} & \textbf{Accuracy} & \textbf{Accuracy} & \textbf{Accuracy}\\
    \midrule
    \textbf{LSTM} & \textbf{C} & 100 epochs &
\SI{82.4136035436025}{\percent}&	\SI{79.300310808525}{\percent}&	\SI{64.3866752650886}{\percent}&	\SI{56.8859616665433}{\percent}
\\
    \midrule
    \multirow{3}{*}{\textbf{\BERTforCode{}}} & \multirow{3}{*}{\textbf{C}} & 2 epochs &
\SI{96.896154}{\percent}&	\SI{94.86645}{\percent}&	\SI{91.14235}{\percent}&	\SI{89.411515}{\percent}
\\
    && 10 epochs &
\SI{97.22805}{\percent}&	\SI{95.487165}{\percent}&	\SI{92.334646}{\percent}&	\SI{90.84322}{\percent}
\\
    && 20 epochs &
\SI{97.27086}{\percent}&	\SI{95.39891}{\percent}&	\SI{92.123246}{\percent}&	\SI{90.6091}{\percent}
\\
    \midrule\midrule
    \multirow{3}{*}{\textbf{\BERTforCode{}}} & \multirow{3}{*}{\textbf{H}}& 2 epochs &
\SI{95.631003}{\percent}&	\SI{90.706295}{\percent}&	\SI{83.50254}{\percent}&	\SI{80.76705}{\percent}
\\
    &  & 10 epochs &
\SI{96.06945}{\percent}&	\SI{91.7122}{\percent}&	\SI{85.373914}{\percent}&	\SI{82.90709}{\percent}
\\
    &  & 20 epochs &
\SI{96.139544}{\percent}&	\SI{91.48951}{\percent}&	\SI{84.847236}{\percent}&	\SI{82.30311}{\percent}
\\
    \midrule
    \citet{hellendoorn2020global} & \textbf{H} &  &
&	\SI{81.9}{\percent}&	&	\SI{73.8}{\percent}
\\
    \bottomrule
\end{tabular}
    \caption{\update{Variable-misuse localization and repair task. Comparison of the LSTM+pointer model~\cite{DBLP:journals/corr/abs-1904-01720} to our fine-tuned \BERTforCode{}+pointer model. We also show results on the test data by \citet{hellendoorn2020global} computed by us and reported by the authors in their Table~1. In the Test Data column, \textbf{C} means our \BERTforCode test dataset, and \textbf{H} means the test dataset used  by \citet{hellendoorn2020global}.}}
    \label{tab:varmisuse-pointer}
\end{table*}

\subsection{The Effects of Context}
\label{sec:SmallerLengths}

Context size is especially useful in code tasks, given that some relevant information may lie many ``sentences'' away from its locus of interest. Here we study how reducing the context length (i.e., the length of the examples used to pre-train and fine-tune) affects performance. We produce data with shorter example lengths, by first pre-training a model on a given example length, and then fine-tuning that model on the corresponding task with examples of that same example length.\footnote{Note that we did not attempt to, say, pre-train on length \num{1024} and then fine-tune that model on length \num{256}-examples, which may also be a practical scenario.} Table~\ref{tab:SmallerLengths} shows the results.

\update{Although context seems to be important to most tasks, the Function Docstring task paradoxically improves with less context. This may be because the task primarily depends on comparison between the docstring and the function signature, and including more context dilutes the model's focus.}


\update{For comparison, we also evaluated the BiLSTM model on varying example lengths for the Variable-Misuse task with CBOW and negative sampling (last column of Table~\ref{tab:SmallerLengths}). More context does seem to benefit the BiLSTM Variable-Misuse classifier as well. However, the improvement offered by \BERTforCode with increasing context is significantly greater.
}

\subsection{{Evaluation on a Multi-Headed Pointer Task}}
\label{sec:varmisuse-pointer}

We now discuss the results of fine-tuning \BERTforCode to predict the localization and repair pointers for the variable-misuse task.
For this task, we implement the multi-headed pointer model from~\citet{DBLP:journals/corr/abs-1904-01720} on top of \BERTforCode.
The baseline consists of the same pointer model on a unidirectional LSTM as used by~\citet{DBLP:journals/corr/abs-1904-01720}.
We refer to these models as \BERTforCode{}+pointer and LSTM+pointer, respectively.
Due to limitations of space, we omit the details of the pointer model and refer the reader to the above paper.
However, the two implementations are identical above the sequence encoding layer;
the difference is the BERT encoder versus an LSTM encoder.
As reported in Section~4 of that work, to enable comparison with an enumerative approach, the evaluation was performed only on 12K test examples.
Instead, here we report the numbers on all \update{378K} of our test examples for both models.

We trained the baseline model for \num{100} epochs and fine-tuned \BERTforCode for \num{2}, \num{10}, and \num{20} epochs. Table~\ref{tab:varmisuse-pointer} gives the results along the same metrics as~\citet{DBLP:journals/corr/abs-1904-01720}. The metrics are defined as follows: 1) True Positive is the percentage of bug-free functions classified as bug-free. 2) Classification Accuracy is the percentage of correctly classified examples (between bug-free and buggy). 3) Localization Accuracy is the percentage of buggy examples for which the localization pointer correctly identifies the bug location. 4) Localization+Repair Accuracy is the percentage of buggy examples for which both the localization and repair pointers make correct predictions. \update{As seen from Table~\ref{tab:varmisuse-pointer} (top 4 rows), \BERTforCode{}+pointer outperforms  LSTM+pointer consistently across all the metrics, and even within 2 and 10 epochs.}

More recently, \citet{hellendoorn2020global} evaluated hybrid models for the same task, combining graph neural networks, Transformers, and RNNs, and greatly improving prior results.
To compare, we obtained the same test dataset from the authors, and evaluated our \BERTforCode fine-tuned model on it. The last four rows of Table~\ref{tab:varmisuse-pointer} show our results and the results reported in that  work.
Interestingly, the models by~\citet{hellendoorn2020global} make use of richer input representations, including syntax, data flow, and control flow. \update{Nevertheless, \BERTforCode outperforms them while using only a lexical representation of the input program.}

\section{Conclusions {and Future Work}}

We present the first attempt at pre-trained contextual embedding of source code by training a BERT model, called \BERTforCode, which we fine-tuned on five {classification} tasks, and compared against BiLSTM with Word2Vec embeddings and Transformer models. {As a more challenging task, we also evaluated \BERTforCode on a multi-headed pointer prediction task.} \BERTforCode outperformed the baseline models consistently. We evaluated \BERTforCode with less data and fewer epochs, highlighting the benefits of pre-training on a massive code corpus. 

We use only source-code tokens and leave it to the underlying Transformer model to infer any structural interactions between them through self-attention. Prior work~\citep{graphsiclr2018, hellendoorn2020global} has argued for explicitly using structural program information (e.g., control flow and data flow). It is an interesting avenue of future work to incorporate such information in pre-training using relation-aware Transformers~\citep{shaw2018self}. However, our improved results in comparison to \citet{hellendoorn2020global} show that \BERTforCode is a simple yet powerful technique and provides a strong baseline for future work on source-code representations.

While surpassing the accuracies achieved by \BERTforCode with newer models and pre-training/fine-tuning methods would be a natural extension to this work, we also envision other follow-up work.
There is increasing interest in developing pre-training methods that can produce smaller models more efficiently and that trade-off accuracy for reduced model size. Further, our benchmark could be valuable to techniques that explore other program representations (e.g., trees and graphs), in multi-task learning, and to develop related tasks such as program synthesis.

\section*{Acknowledgements}

We are indebted to Daniel Tarlow for his guidance and generous advice throughout the development of this work. Our work has also improved thanks to feedback, use cases, helpful libraries, and proofs of concept offered by David Bieber, Vincent Hellendoorn, Ben Lerner, Hyeontaek Lim, Rishabh Singh, Charles Sutton, and Manushree Vijayvergiya. Finally, we are grateful to the anonymous reviewers, who gave useful, constructive comments and helped us improve our presentation and results.

\bibliography{combined}
\bibliographystyle{icml2020}

\clearpage

\appendix
\section{Open-Sourced Artifacts}

We release data and some source-code utilities at \url{https://github.com/google-research/google-research/tree/master/cubert}. The repository contains the following:
\begin{description}
    \item[GitHub Manifest] A list of all the file versions we included into our pre-training corpus, after removing files similar to the fine-tuning corpus\footnote{\url{https://github.com/google-research-datasets/eth_py150_open}}, and after deduplication. The manifest can be used to retrieve the file contents from GitHub or Google's BigQuery. This dataset was retrieved from Google's BigQuery on June 21, 2020.
    \item[Vocabulary] Our subword vocabulary, computed from the pre-training corpus.
    \item[Pre-trained Models] Pre-trained models on the pre-training corpus, after 1 and 2 epochs, for examples of length 512, and the BERT Large architecture.
    \item[Task Datasets] Datasets containing training, validation, and testing examples for each of the 6 tasks. For the classification tasks, we provide original source code and classification labels. For the localization and repair task, we provide subtokenized code, and masks specifying the targets.
    \item[Fine-tuned Models] Fine-tuned models for the 6 tasks. Fine-tuning was done on the 1-epoch pre-trained model. For each classification task, we provide the checkpoint with highest validation accuracy; for the localization and repair task, we provide the checkpoint with highest localization and repair accuracy. These are the checkpoints we used to evaluate on our test datasets, and to compute the numbers in the main paper.
    \item[Code-encoding Library] We provide code for tokenizing Python code, and for producing inputs to \BERTforCode{}'s pre-training and fine-tuning models.
    \item[Localization-and-repair Fine-tuning Model] We provide a library for constructing the localization-and-repair model, on top of \BERTforCode{}'s encoder layers. For the classification tasks, the model is identical to that of BERT's classification fine-tuning model.
\end{description}
Please see the \texttt{README} for details, file encoding and schema, and terms of use.

\section{{Data Preparation for Fine-Tuning Tasks}}
\label{sec:data-prep}

\subsection{{Label Frequencies}}
All four of our binary-classification fine-tuning tasks had an equal number of buggy and bug-free examples.
The Exception task, which is a multi-class classification task, had a different number of examples per class (i.e., exception types). For the Exception task, we show the breakdown of example counts per label for our fine-tuning dataset splits in Table~\ref{tab:exceptionclasses}.

\begin{table*}
\centering
\begin{tabular}{|l|r|r|rrr|}
\hline
\multicolumn{1}{|c|}{\multirow{2}{*}{\textbf{Exception Type}}} & \multicolumn{1}{c|}{\multirow{2}{*}{\textbf{Test}}} & \multicolumn{1}{c|}{\multirow{2}{*}{\textbf{Validation}}} & \multicolumn{3}{c|}{\textbf{Train}}                                              \\
\multicolumn{1}{|c|}{}                                         & \multicolumn{1}{c|}{}                               & \multicolumn{1}{c|}{}                                     & \multicolumn{1}{l}{100\%} & \multicolumn{1}{l}{66\%} & \multicolumn{1}{l|}{33\%} \\ \hline
\verb+ASSERTION_ERROR+&	\num{155}&	\num{29}&	\num{323}&	\num{189}&	\num{86}\\
\verb+ATTRIBUTE_ERROR+&	\num{1372}&	\num{274}&	\num{2444}&	\num{1599}&	\num{834}\\
\verb+DOES_NOT_EXIST+&	\num{7}&	\num{2}&	\num{3}&	\num{3}&	\num{2}\\
\verb+HTTP_ERROR+&	\num{55}&	\num{9}&	\num{104}&	\num{78}&	\num{38}\\
\verb+IMPORT_ERROR+&	\num{690}&	\num{170}&	\num{1180}&	\num{750}&	\num{363}\\
\verb+INDEX_ERROR+&	\num{586}&	\num{139}&	\num{1035}&	\num{684}&	\num{346}\\
\verb+IO_ERROR+&	\num{721}&	\num{136}&	\num{1318}&	\num{881}&	\num{427}\\
\verb+KEY_ERROR+&	\num{1926}&	\num{362}&	\num{3384}&	\num{2272}&	\num{1112}\\
\verb+KEYBOARD_INTERRUPT+&	\num{232}&	\num{58}&	\num{509}&	\num{336}&	\num{166}\\
\verb+NAME_ERROR+&	\num{78}&	\num{19}&	\num{166}&	\num{117}&	\num{60}\\
\verb+NOT_IMPLEMENTED_ERROR+&	\num{119}&	\num{24}&	\num{206}&	\num{127}&	\num{72}\\
\verb+OBJECT_DOES_NOT_EXIST+&	\num{95}&	\num{16}&	\num{197}&	\num{142}&	\num{71}\\
\verb+OS_ERROR+&	\num{779}&	\num{131}&	\num{1396}&	\num{901}&	\num{459}\\
\verb+RUNTIME_ERROR+&	\num{107}&	\num{34}&	\num{247}&	\num{159}&	\num{80}\\
\verb+STOP_ITERATION+&	\num{270}&	\num{61}&	\num{432}&	\num{284}&	\num{131}\\
\verb+SYSTEM_EXIT+&	\num{105}&	\num{16}&	\num{200}&	\num{120}&	\num{52}\\
\verb+TYPE_ERROR+&	\num{809}&	\num{156}&	\num{1564}&	\num{1038}&	\num{531}\\
\verb+UNICODE_DECODE_ERROR+&	\num{134}&	\num{21}&	\num{196}&	\num{135}&	\num{63}\\
\verb+VALIDATION_ERROR+&	\num{92}&	\num{16}&	\num{159}&	\num{96}&	\num{39}\\
\verb+VALUE_ERROR+&	\num{2016}&	\num{415}&	\num{3417}&	\num{2232}&	\num{1117}\\
\hline
\end{tabular}

\caption{Example counts per class for the Exception Type task, broken down into the dataset splits. We show separately the 100\% train dataset, as well as its 33\% and 66\% subsamples used in the ablations.}
\label{tab:exceptionclasses}
\end{table*}



\subsection{{Fine-Tuning Task Datasets}}
In this section, we describe in detail how we produced our fine-tuning datasets (Section~3.4 of the main paper).

A common primitive in all our data generation is splitting a Python module into functions. We do this by parsing the Python file and identifying function definitions in the Abstract Syntax Tree that have no other function definition between themselves and the root of the tree. The resulting functions include functions defined at module scope, but also methods of classes and subclasses. Not included are functions defined within other function and method bodies, or methods of classes that are, themselves, defined within other function or method bodies.

We do not filter functions by length, although task-specific data generation may filter out some functions (see below). When generating examples for a fixed-length pre-training or fine-tuning model, we prune all examples to the maximum target sequence length (in this paper we consider \num{128}, \num{256}, \num{512}, and \num{1024} subtokenized sequence lengths). Note that if a synthetically generated buggy/bug-free example pair differs only at a location beyond the target length (say on the \num{2000}-th subtoken), we still retain both examples. For instance, for the Variable-Misuse Localization and Repair task, we retain both buggy and bug-free examples, even if the error and/or repair locations lie beyond the end of the maximum target length. During evaluation, if the error or repair locations fall beyond the length limit of the example, we count the example as a model failure.

\subsubsection{Reproducible Data Generation}
We make pseudorandom choices at various stages in fine-tuning data generation. It was important to design a pseudorandomness mechanism that gave (a) reproducible data generation, (b) non-deterministic choices drawn from the uniform distribution, and (c) order independence. Order independence is important because our data generation is done in a distributed fashion (using Apache Beam), so different pseudorandom number generator state machines are used by each distributed worker.

Pseudorandomness is computed based on an experiment-wide seed, but is independent of the order in which examples are generated. Specifically, to make a pseudorandom choice about a function, we hash (using MD5) the seed and the function data (its source code and metadata about its provenance), and use the resulting hash as a uniform pseudorandom value from the function, for whatever needs the data generator has (e.g., in choosing one of multiple choices). In that way, the same function will always result in the same choices given a seed, regardless of the order in which each function is processed, thereby ensuring reproducible dataset generation.

To select among multiple choices, we hash the function's pseudorandom value along with all choices (sorted in a canonical order) and use the digest to compute an index within the list of choices. Note that given two choices over different candidates but for the same function, independent decisions will be drawn. We also use such order-independent pseudorandomness when subsampling datasets (e.g., to generate the validation datasets). In those cases, we hash a sample with the seed, as above, and turn the resulting digest into a pseudorandom number in $[0, 1]$, which can be used to decide given a target sampling rate.

\subsubsection{Variable-Misuse Classification}
\label{sec:varmisuseclassificationdata}
A variable use is any mention of a variable in a load scope. This includes a variable that appears in the right-hand side of an assignment, or a field dereference. We regard as \emph{defined} all variables mentioned either in the formal arguments of a function definition, or on the left-hand side of an assignment. We do not include in our defined variables those declared in module scope (i.e., globals).

To decide whether to generate examples from a function, we parse it, and collect all variable-use locations, and all defined variables, as described above. We discard the function if it has no variable uses, or if it defines fewer than two variables; this is necessary, since if there is only one variable defined, the model has no choice to make but the default one. We also discard the function if it has more than \num{50} defined variables; such functions are few, and tend to be auto-generated. For any function that we do not discard, i.e., an \emph{eligible} function, we generate a buggy and a bug-free example, as described next.

To generate a buggy example from an eligible function, we choose one variable use pseudorandomly (see above how multiple-choice decisions are done), and replace its current occupant with a different pseudorandomly-chosen variable defined in the function (with a separate multiple-choice decision).

Note that in the work by \citet{DBLP:journals/corr/abs-1904-01720}, a buggy and bug-free example pair was generated for \emph{every} variable use in an eligible function. In the work by \citet{hellendoorn2020global}, a buggy and bug-free example pair was generated for \emph{up to three} variable uses in an eligible function, i.e., some functions with one use would result in one example pair, whereas functions with many variable uses would result in three example pairs. In contrast, our work produces \emph{exactly one} example pair for every eligible function. Eligibility was defined identically in all three projects.

\subsubsection{Wrong Binary Operator}
This task considers both commutative and non-commutative binary operators (unlike the Swapped-Argument Classification task). See Table~\ref{tab:binaryoperators} for the full list, and note that we have excluded relatively infrequent operators, e.g., the Python integer division operator \texttt{//}.

\begin{table}
\centering
\begin{tabular}{|c|c|c|}
\hline
                    & \textbf{Commutative} & \textbf{Non-Commutative}                                   \\ \hline
\textbf{Arithmetic} & +, *                 & -, /, \%                                                   \\
\textbf{Comparison} & ==, !=, is, is not   & \textless{}, \textless{}=, \textgreater{}, \textgreater{}= \\
\textbf{Membership} &                      & in, not in                                                 \\
\textbf{Boolean}    & and, or              &                                                            \\ \hline
\end{tabular}
\caption{Binary operators.}
\label{tab:binaryoperators}
\end{table}

If a function has no binary operators, it is discarded. Otherwise, it is used to generate a bug-free example, and a single buggy example as follows: one of the operators is chosen pseudorandomly (as described above), and a different operator chosen to replace it from the same row of  Table~\ref{tab:binaryoperators}. So, for instance, a buggy example would only swap \texttt{==} with \texttt{is}, but not with \texttt{not in}, which would not type-check if we performed static type inference on Python.

We take appropriate care to ensure the code parses after a bug is introduced. For instance, if we swap the operator in the expression \texttt{1==2} with \texttt{is}, we ensure that there is space between the tokens (i.e., \texttt{1 is 2} rather than the incorrect \texttt{1is2}), even though the space  was not needed before.

\subsubsection{Swapped Operand}
Since this task targets swapping the arguments of binary operators, we only consider non-commutative operators from Table~\ref{tab:binaryoperators}.

Functions without eligible operators are discarded, and the choice of the operator to mutate in a function, as well as the choice of buggy operator to use, are done as above, but limiting choices only to non-commutative operators.

To avoid complications due to format changes, we only consider expressions that fit in a single line (in contrast to the Wrong Binary Operator Classification task). We also do not consider expressions that look the same after swapping (e.g., \texttt{a - a}).

\subsubsection{Function-Docstring Mismatch}
In Python, a function docstring is a string literal that directly follows the function signature and precedes the main function body. Whereas in other common programming languages, the function documentation is a comment, in Python it is an actual, semantically meaningful string literal.

We discard functions that have no docstring from this dataset, or functions that have an empty docstring. We split the rest into the function definition without the docstring, and the docstring summary (i.e., the first line of text from its docstring), discarding the rest of the docstring.

We create bug-free examples by pairing a function with its own docstring summary.

To create buggy examples, we pair every function with another function's docstring summary, according to a global pseudorandom permutation of all functions: for all $i$, we combine the $i$-th function (without its docstring) with the $P_i$-th function's docstring summary, where $P$ is a pseudorandom permutation, under a given seed. We discard pairings in which $i == P[i]$, but for the seeds we chose, no such pathological permuted pairings occurred.

\subsubsection{Exception Type}
Note that, unlike all other tasks, this task has no notion of buggy or bug-free examples.

We discard functions that do not have any \texttt{except} clauses in them.

For the rest, we collect all locations holding exception types within 
\texttt{except} clauses, and choose one of those locations to query the model for classification. Note that a single \texttt{except} clause may hold a comma-separated list of exception types, and the same type may appear in multiple locations within a function. Once a location is chosen, we replace it with a special \texttt{\_\_HOLE\_\_} token, and create a classification example that pairs the function (with the masked exception location) with the true label (the removed exception type).

The count of examples per exception type can be found in Table~\ref{tab:exceptionclasses}.

\subsubsection{{Variable Misuse Localization and Repair}}
The dataset for this task is identical to that for the Variable-Misuse Classification task (Section~\ref{sec:varmisuseclassificationdata}). However, unlike the classification task, examples contain more features relevant to localization and repair. Specifically, in addition to the token sequence describing the program, we also extract a number of boolean input masks:
\begin{itemize}
    \item A \emph{candidates} mask, which marks as True all tokens holding a variable, which can therefore be either the location of a bug, or the location of a repair. The first position is always a candidate, since it may be used to indicate a bug-free program.
    \item A \emph{targets} mask, which marks as True all tokens holding the correct variable, for buggy examples. Note that the correct variable may appear in multiple locations in a function, therefore this mask may have multiple True positions. Bug-free examples have an all-False targets mask.
    \item An \emph{error-location} mask, which marks as True the location where the bug occurs (for buggy examples) or the first location (for bug-free examples).
\end{itemize}
All the masks mark as True some of the locations that hold variables. Because many variables are subtokenized into multiple tokens, if a variable is to be marked as True in the corresponding mask, we only mark as True its first subtoken, keeping trailing subtokens as False.

\section{Attention Visualizations}
\label{sec:visualizations}

\begin{figure*}
    \centering
    \begin{tabular}{c}
    \includegraphics[width=.8\textwidth]{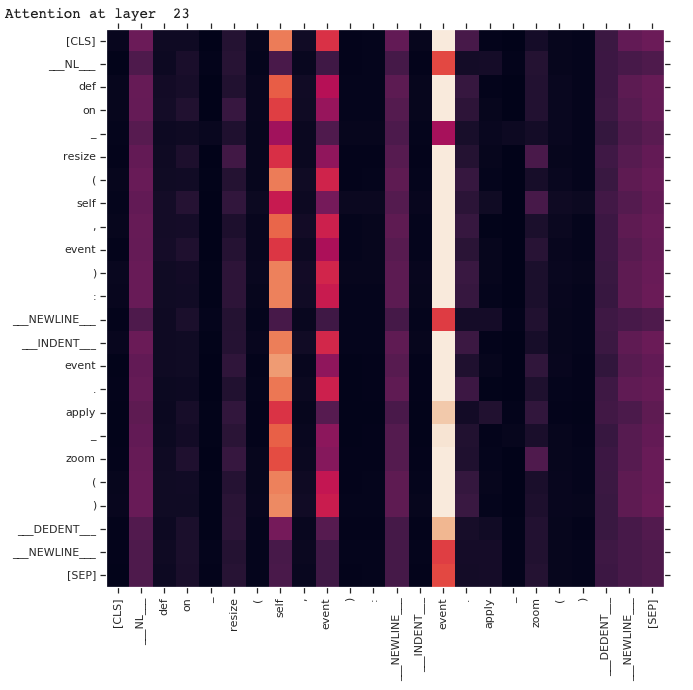}\\
    \begin{lstlisting}[linewidth=0.5\textwidth]
def on_resize(self, event):
  event.apply_zoom()
    \end{lstlisting} 
    \end{tabular}
    \caption{Variable Misuse Example. In the code snippet, `{\tt event.apply\_zoom}' should actually be `{\tt self.apply\_zoom}'. The CuBERT variable-misuse model correctly predicts that the code has an error. As seen from the attention map, the query tokens are attending to the second occurrence of the `{\tt event}' token in the snippet, which corresponds to the incorrect variable usage.}
    \label{fig:misuse_visualization}
\end{figure*}

\begin{figure*}
    \centering
    \begin{tabular}{c}
    \includegraphics[height=.8\textwidth]{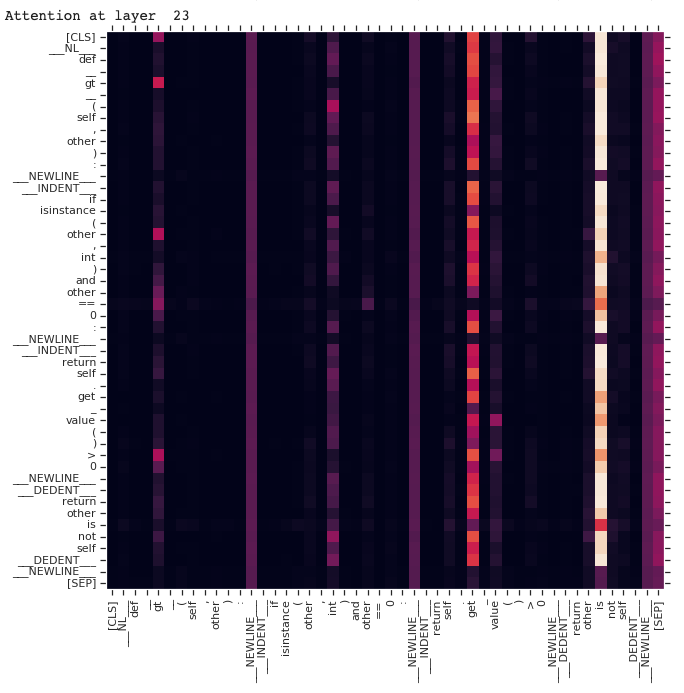}\\
    \begin{lstlisting}[linewidth=0.6\textwidth]
def__gt__(self,other):
  if isinstance(other,int)and other==0:
    return self.get_value()>0
  return other is not self
      \end{lstlisting} 
    \end{tabular}
    \caption{Wrong Operator Example. In this code snippet, `\texttt{other is not self}' should actually be `\texttt{other < self}'. The CuBERT wrong-binary-operator model correctly predicts that the code snippet has an error. As seen from the attention map, the query tokens are all attending to the incorrect operator `\texttt{is}'.}
    \label{fig:operator_visualization}
\end{figure*}

\begin{figure*}
    \centering
    \begin{tabular}{c}
    \includegraphics[height=.8\textwidth]{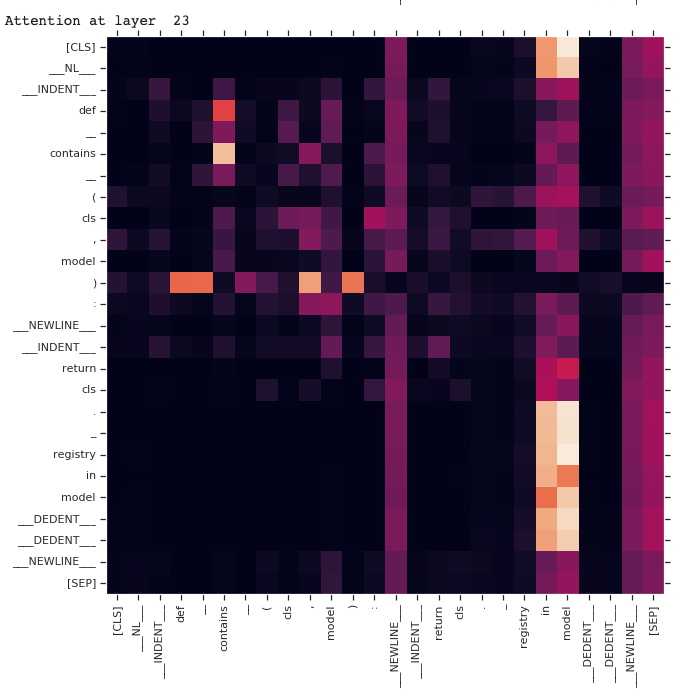}\\
    \begin{lstlisting}[linewidth=0.5\textwidth]
def__contains__(cls,model):
  return cls._registry in model
      \end{lstlisting} 
    \end{tabular}
    \caption{Swapped Operand Example. In this code snippet, the return statement should be `\texttt{model in cls.\_registry}'. The 
    swapped-operand model correctly predicts that the code snippet has an error. The query tokens are paying substantial attention to `\texttt{in}' and the second occurrence of `\texttt{model}' in the snippet.}
    \label{fig:operand_visualization}
\end{figure*}

\begin{figure*}
    \centering
    \begin{tabular}{c}
    \includegraphics[height=.8\textwidth]{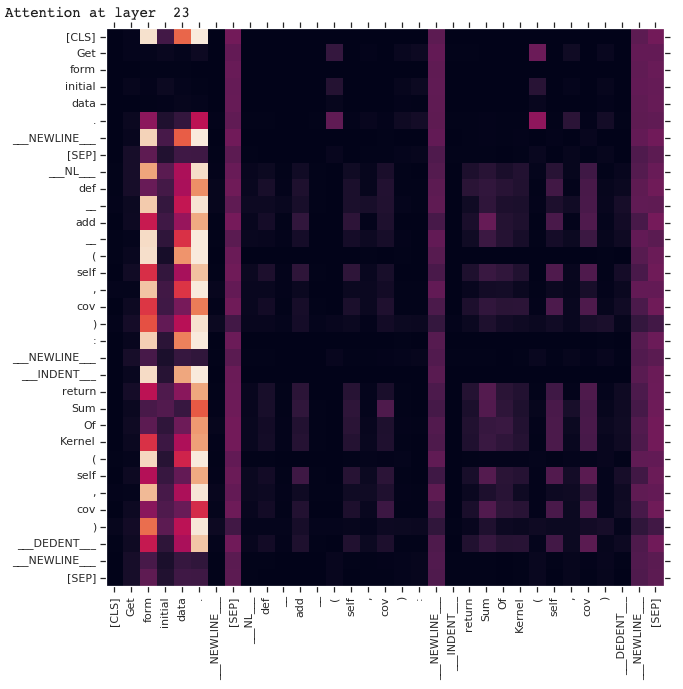}\\
    \begin{lstlisting}[linewidth=0.6\textwidth]
Docstring: 'Get form initial data.'
Function:
def__add__(self,cov):
  return SumOfKernel(self,cov)
      \end{lstlisting} 
    \end{tabular}
    \caption{Function Docstring Example. The CuBERT function-docstring model correctly predicts that the docstring is wrong for this code snippet. Note that most of the query tokens are attending to the tokens in the docstring.}
    \label{fig:docstring_visualization}
\end{figure*}

\begin{figure*}
    \centering
    \begin{tabular}{c}
    \includegraphics[height=.8\textwidth]{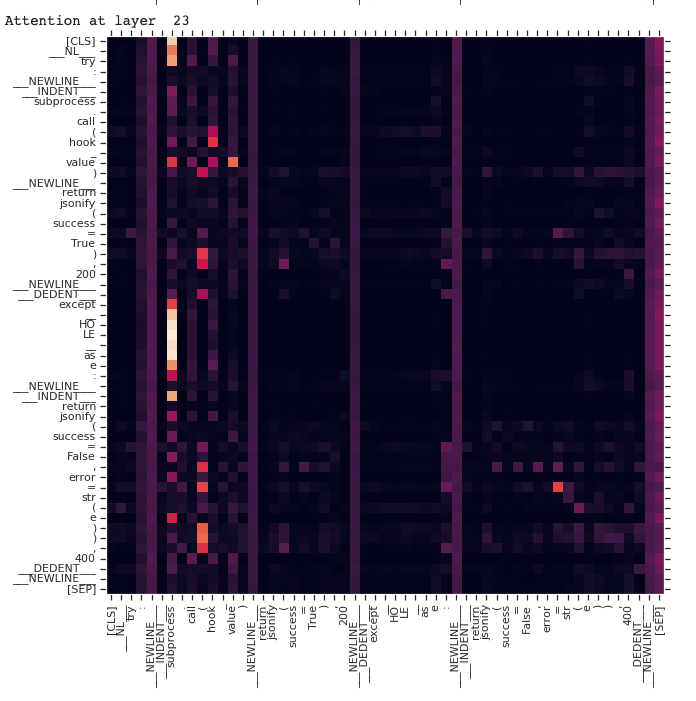}\\
    \begin{lstlisting}[linewidth=0.6\textwidth]
try:
  subprocess.call(hook_value)
  return jsonify(success=True), 200
except __HOLE__ as e:
  return jsonify(success=False,
    error=str(e)), 400
      \end{lstlisting} 
    \end{tabular}
    \caption{Exception Classification Example. For this code snippet, the CuBERT exception-classification model correctly predicts `{\tt \_\_HOLE\_\_}' as `{\tt OSError}'. The model's attention matrix also shows that `{\tt \_\_HOLE\_\_}' is attending to `{\tt subprocess}', which is indicative of an OS-related error.}
    \label{fig:exception_visualization}
\end{figure*}

In this section, we provide sample code snippets used to test the different classification tasks. Further, Figures \ref{fig:misuse_visualization}--\ref{fig:exception_visualization} show visualizations of the attention matrix of the last layer of the fine-tuned CuBERT model~\citep{Coenen2019VisualizingAM} for the code snippets. In the visualization, the Y-axis shows the query tokens and X-axis shows the tokens being attended to. The attention weight between a pair of tokens is the maximum of the weights assigned by the multi-head attention mechanism. The color changes from dark to light as weight changes from 0 to 1.

\end{document}